# Modeling of the ionosphere response on the earthquake preparation

## M.I. Karpov, O.V. Zolotov, A.A. Namgaladze
*Polytechnic Faculty of MSTU, Physics Department*

**Abstract.** Seismo-ionosphere coupling processes have been investigated considering the GPS observed anomalous ionospheric Total Electron Content (TEC) variations before strong earthquakes as their precursors. The numerical simulations' results of the TEC response on the vertical electric currents flowing between the Earth and ionosphere during the earthquake (EQ) preparation time have been performed. Model experiments have been carried out using the Upper Atmosphere Model. The following currents' parameters were varied in: (i) direction (to or from the ionosphere); (ii) latitudinal zone of the sources' (EQ epicenters) location; (iii) currents' configuration: (1) grid nodes with "straight" currents were surrounded by "border" grid points with currents of opposite direction ("return" currents); (2) the "return" currents were spread out over the globe; (3) without "return" currents. Numerical simulations have shown that electric currents with density of $4 \cdot 10^{-8}$ A/m$^2$ over the area of about 200 km in longitude and 2500 km in latitude produce both positive and negative TEC disturbances with magnitude up to 35 % in agreement with GPS TEC observations before EQs.

**Аннотация.** Исследованы проявления сейсмо-ионосферных связей – аномальных возмущений полного электронного содержания (ПЭС) ионосферы перед землетрясениями как их предвестников. Представлены результаты численного моделирования полного электронного содержания (ПЭС) в зависимости от вертикальных электрических токов между Землей и ионосферой перед землетрясениями. Модельные эксперименты проводились с помощью модели верхней атмосферы UAM. Варьировались следующие параметры токов: направление, широтное расположение источников (эпицентров землетрясений), конфигурации токов: узлы с "прямыми" токами были окружены пограничными узлами с токами противоположного направления ("возвратными" токами); "возвратные" токи были распределены по всей земной поверхности; без "возвратных" токов. Расчеты показали, что токи плотностью $4 \cdot 10^{-8}$ A/м$^2$ над областью около 200 км по долготе и 2500 км по широте вызывают положительные и отрицательные возмущения ПЭС магнитудой до 35 % и более, что соответствует GPS наблюдениям перед землетрясениями.

**Ключевые слова:** полное электронное содержание, сейсмический электрический ток, предвестники землетрясений, ионосфера, UAM, моделирование

## 1. Introduction

Seismogenic electric fields generated near the tectonic faults are important elements of the global electric circuit (e.g., *Harrison et al.*, 2010; *Pulinets, Boyarchuk*, 2004, and references therein; *Freund et al.*, 2009; *Pulinets, Ouzounov*, 2011; *Sorokin et al.*, 2005; 2006; 2007). Radioactive radon emanation (*Pulinets, Boyarchuk*, 2004) and "positive holes" mechanism (*Freund et al.*, 2009) both are acting as air ionization sources over the faults. According to *Freund* (2009) stressed rocks could generate electric current with density of 0.5-1.25 μA/m$^2$. *Sorokin et al.* (2005; 2006; 2007) calculated that such currents within an area about 200 km in radius are able to create ionosphere electric field disturbances of several mV/m. Such electric fields in turn may disturb the ionosphere F-2 electron density and total electron content (TEC) up to 50 % and more (*Namgaladze et al.*, 2007; 2009a; 2009b).

Many other studies, observations and experiments demonstrate that vertical electric currents with such densities exist in nature. For example, *Krider* and *Musser* (1982) used data of time variations in thunderstorm electric fields, both aloft and at the ground, provided by a large field mill network. They estimated that the average Maxwell current density under a small Florida thunderstorm ranged from 1 to 4 nA/m$^2$ for a short period of time. The air conductivity and the vertical electric field were measured over thunderstorms from U-2 airplane by *Blakeslee, Christian* (1989). It was shown that the current with density ranging from 0 to 20 nA/m$^2$ flows upward from thunderstorm areas. *Le Mouel et al.* (2010) observed transient electric potential variations in a standing poplar tree equipped with electrodes along its height. It was obtained that current with typical value of 2.5 μA flows through the average cross section of the trunk of 0.5 m$^2$. It corresponds to current density of 5 μA/m$^2$.

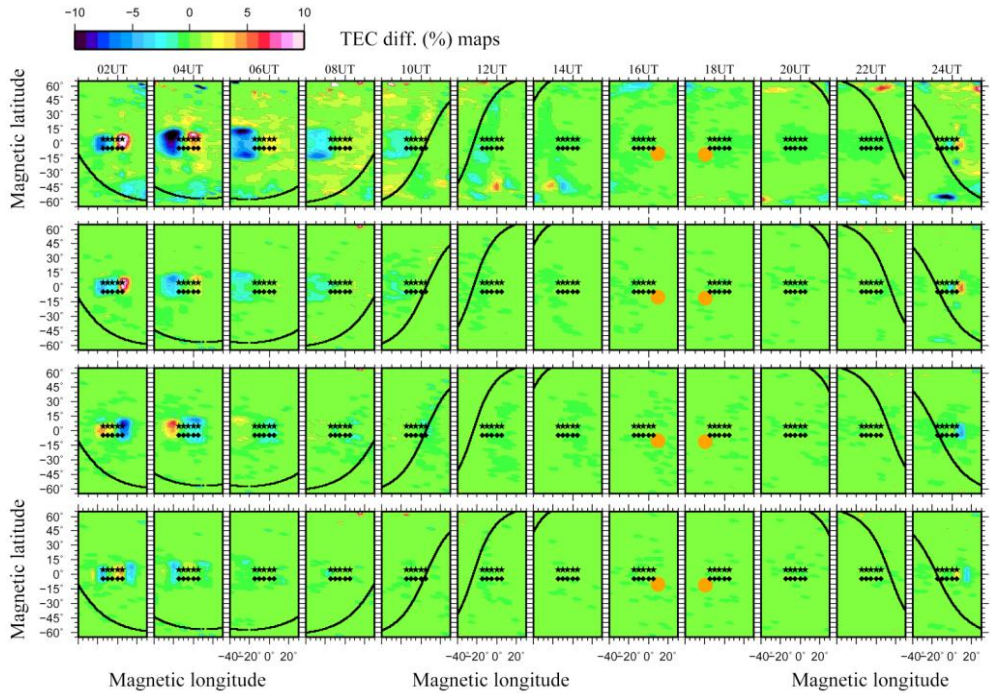

Fig. 1. Magnetic maps of model TEC deviations (%) relative to the quiet conditions for "straight" currents located at 5° lat. (from top to bottom): "return" currents spread out over the globe except points with "straight" currents; without "return" currents; the same as at top but currents flow from the ionosphere to the Earth; "return" currents are located at the grid nodes nearest to the "straight" currents. Stars – grid points with "straight" currents. Diamonds – the magnetically conjugated points. Orange circle – the subsolar point. Black curves – terminator lines

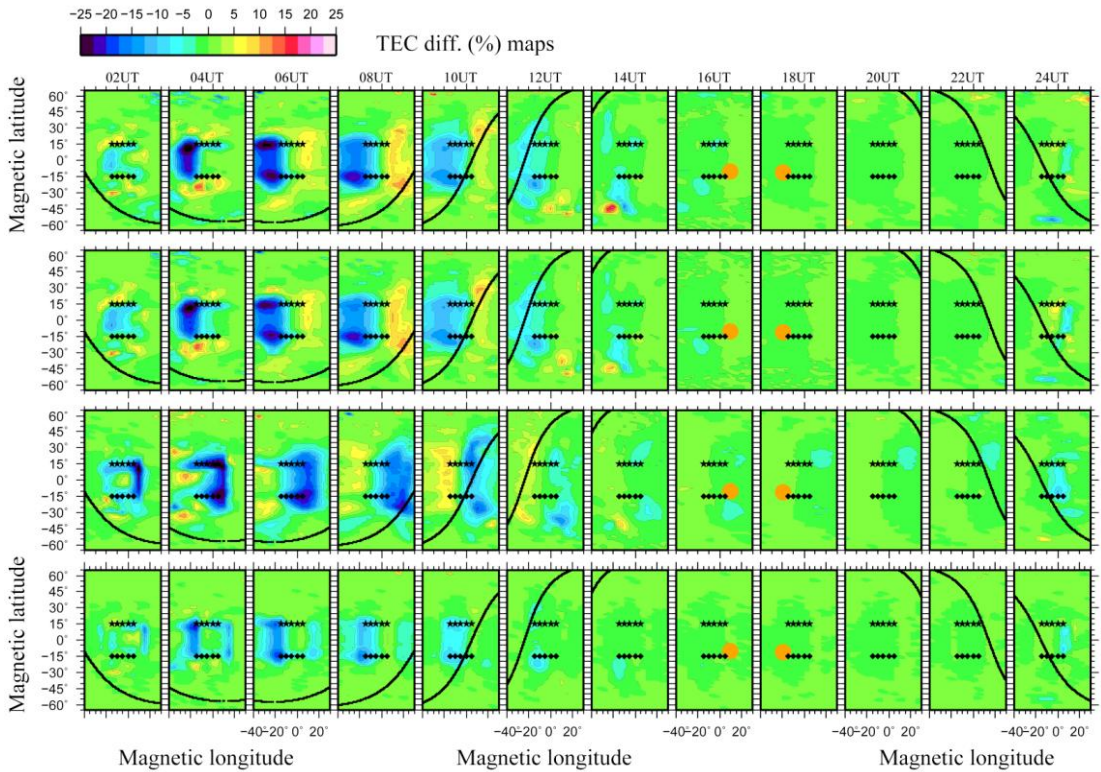

Fig. 2. Magnetic maps of model TEC deviations (%) relative to the quiet conditions
for "straight" currents located at 15° lat.

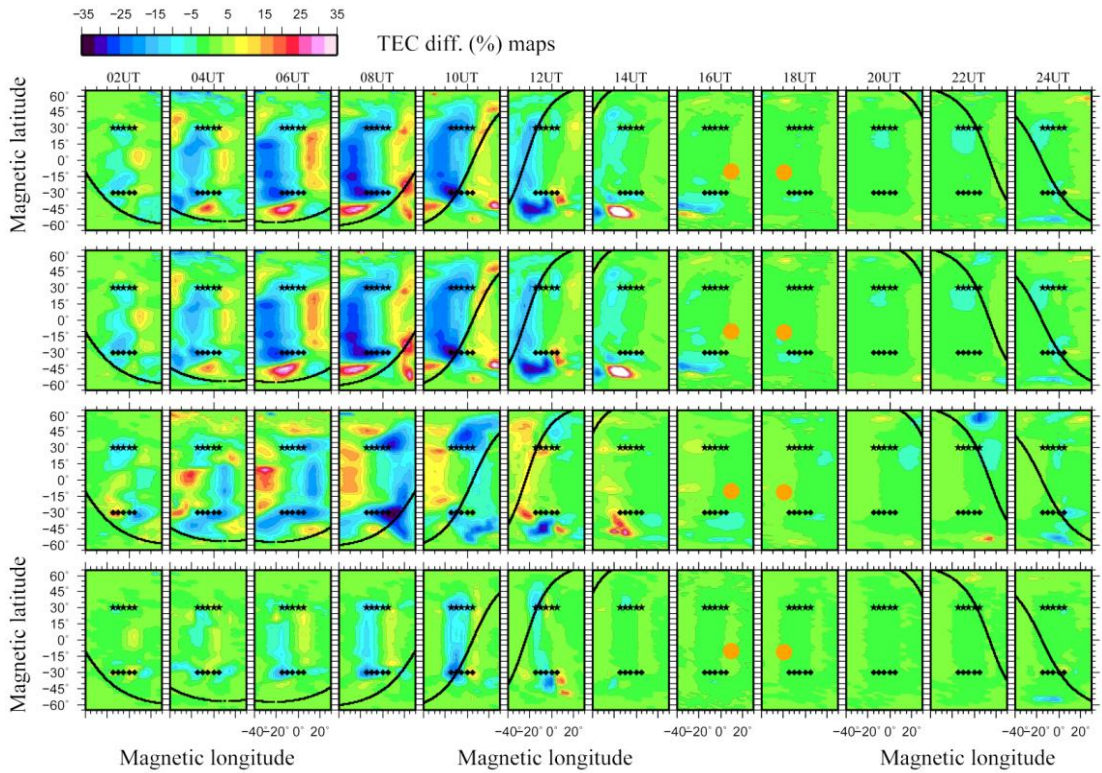

Fig. 3. Magnetic maps of model TEC deviations (%) relative to the quiet conditions for "straight" currents located at 30° lat.

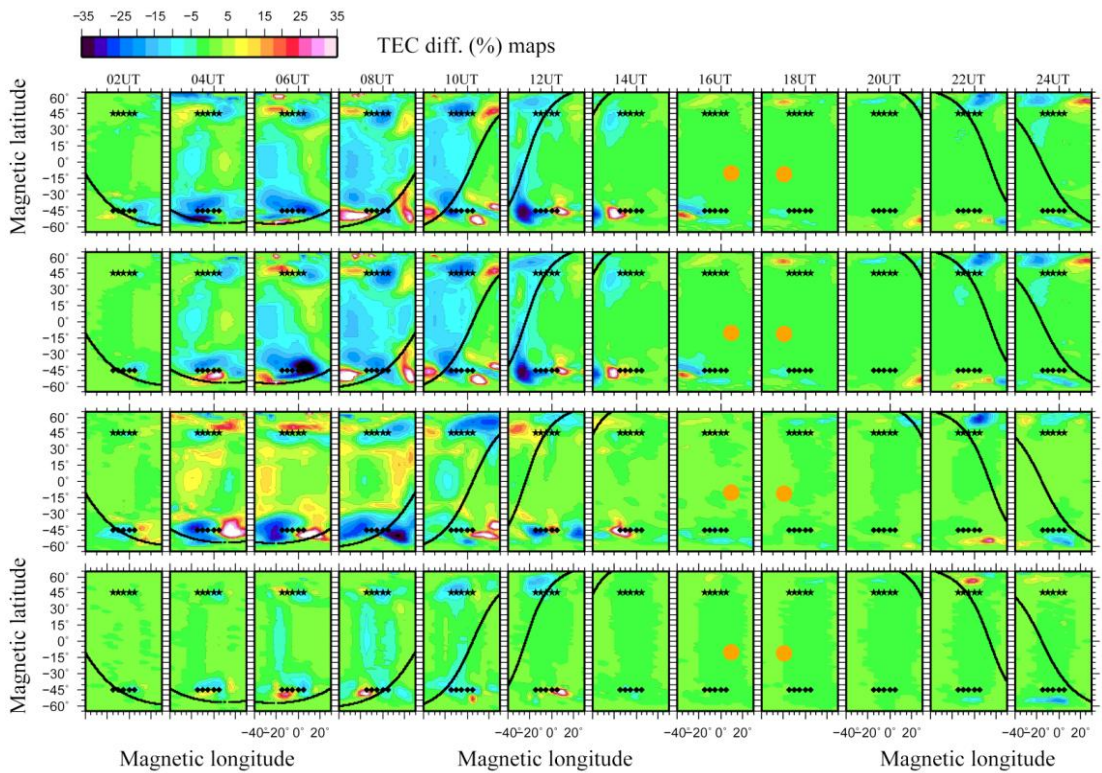

Fig. 4. Magnetic maps of model TEC deviations (%) relative to the quiet conditions for "straight" currents located at 45° lat.

Previous numerical simulations (*Namgaladze*, 2010; *Namgaladze et al.*, 2011; *Namgaladze, Zolotov*, 2011) showed that point electric current sources larger than $10^{-6}$ A/m$^2$ generated too strong TEC disturbances. Similar point sources but with densities less than $10^{-9}$-$10^{-8}$ A/m$^2$ generated TEC disturbances not exceeding 10-20 %. Those values were smaller than ones usually taken into consideration in pre-earthquake TEC modifications studies. The vertical electric current with density of about $2 \cdot 10^{-8}$ A/m2 flowing between the fault and the ionosphere set at the area of about ~200 km × ~4000 km may create electric fields generating the TEC increases up to ~50 % at night-time as observed before the Haiti Jan. 12, 2010 earthquake. Appearance of the terminator and subsolar point caused depression down to full destruction of electric potential generated by external electric current. TEC disturbances were also modified: "escaped" from terminator to dark-side of the ionosphere with following full destruction, but with time lag relative to the electric potential. Next night the electric potential and corresponding TEC disturbances restored.

The principle goal of this paper is to present the results of numerical simulations of ionosphere electric fields and corresponding TEC disturbances depending on various parameters of electric currents between the Earth and ionosphere: (1) latitudinal position, (2) general spatial orientation of the vertical electric currents' distribution and (3) direction (to or from the ionosphere).

**2. Simulation methods**

Model experiments were carried out using the Upper Atmosphere Model – global numerical model of the Earth's upper atmosphere (thermosphere, ionosphere, plasmasphere and inner magnetosphere). The model calculates main physical parameters of upper atmosphere such as densities, temperatures and velocities of the neutral (O, $O_2$, $N_2$, H) and charged ($O_2^+$, $NO^+$, $O^+$, $H^+$ and electrons) species by numerical integration of the continuity, momentum and heat balance equations as well as the equation of the electric field potential (*Namgaladze et al.*, 1988; 1991; 1998a; 1998b).

To build model difference maps of the TEC we first performed a regular calculation without any additional electric current sources to use those results as quiet background values. Then, an external electric current ("straight" current) flowing between the lower atmosphere and the ionosphere presumably of seismogenic origin was used as a model input for the calculations of the ionospheric field and the corresponding TEC variations. These additional sources of the electric current with the magnitude of $4 \cdot 10^{-8}$ A/m$^2$ were switched on at the lower boundary (80 km) in the UAM electric potential equation, which was solved numerically jointly with all other UAM equations for neutral and ionized gases.

Currents sources were switched on at five numerical grid nodes which correspond to the size of the area about 200 km along the parallel and about 3000 km along the meridian depending on latitude. The latitudinal position (5°, 15°, 30° and 45° mag. lat.) and direction of the currents (to or from the ionosphere) were varied in. The grid nodes with vertical electric currents ("straight" currents) were surrounded by boundary grid points with currents of the opposite direction ("return" currents) so that the total current in the global electric circuit didn't change. In other investigated configuration "return" currents were spread out over the globe except the points with "straight" current. The configuration without "return" currents were studied as well.

**3. Discussion and summary**

According to UAM simulations (Fig. 1-4) sources of the additional electric currents flowing between the Earth and ionosphere presumably of seismogenic origin with current density $4 \cdot 10^{-8}$ A/m$^2$ over the area of about 200 km in latitude and 3000 km in longitude produce both positive and negative TEC disturbances. The size and geographical position of the disturbed area didn't change for a few hours. Approaching of the sunlit ionosphere is followed by disappearance of the disturbances.

The results of the numerical computations are in good agreement with the observations of TEC before strong earthquakes at middle and low latitudes in the spatial scales and amplitude characteristics (*Liu et al.*, 2000; 2006; 2011; *Pulinets, Boyarchuk*, 2004; *Pulinets, Ouzhounov*, 2011; *Zakharenkova*, 2006; 2008; *Zolotov et al.*, 2012).

As Figs. 1-4 show changing the direction of the currents causes mirror reverse of the positive and negative disturbance relatively magnetic meridian crossing the epicenter area and magnetically conjugated area except the case for currents at 45° lat.

Various configurations of the currents have different influence on the TEC disturbances. The configuration without "return" currents creates TEC disturbances very similar to those produced by configuration with "return" currents spread out over the whole globe. The magnitude of the TEC disturbances produced by "return" currents located in boundary grid nodes is lower than magnitudes of the disturbances created by other configurations.

The most powerful magnitudes (35 % and larger) are produced by currents located at 30° and 45° mag. lat. The shift of the vertical electric current from 30° magnetic latitude towards the magnetic equator leads to less strong disturbances by magnitude. Sources for latitudes less than 30° generate pronounced modification of the Appleton anomaly.

It should be noted that similar numerical calculations were performed by (*Kuo et al.*, 2011). The difference is that their currents flow the over area 200 km × 30 km that is much smaller than in our case. "Rock currents" were switched on near ground surface (not at the lower boarder of the ionosphere). Then they used the empirical model of the atmosphere's conductivity modified from some considerations. They simulated only TEC response at the epicenter region and did not consider magnetically conjugated ones. The simulations were local (not for the globe as in our case) with "single" point sources. According to their simulations using SAMI2 model the current density of 0.2-10 $\mu A/m^2$ in the earthquake fault zone can cause TEC variations of up to 2-25 % in the daytime ionosphere and the density of 0.01-1 $\mu A/m^2$ can lead to nighttime TEC variations of 1-30 %. Such huge currents are required due to a too small area of the current generation in their simulation. Beside of this, our and their model results do not contradict each other.